\documentclass{PoS}

\usepackage{subfigure}

\title{Gamma-ray flares from black hole coronae.}

\ShortTitle{Gamma-ray flares from black hole coronae}

\author{Gustavo E. Romero\\
        Instituto Argentino de Radioastronom\'ia (IAR-CONICET)\\
        Facultad de Ciencias Astron\'omicas y Geof\'isicas (FCAG, UNLP)\\
        E-mail: \email{romero@iar-conicet.gov.ar}}

\author{Florencia L. Vieyro\\
        Instituto Argentino de Radioastronom\'ia (IAR-CONICET)\\
        Facultad de Ciencias Astron\'omicas y Geof\'isicas (FCAG, UNLP)\\
        E-mail: \email{fvieyro@iar-conicet.gov.ar}}

\abstract{We present results of a study of non-thermal, time-dependent particle injection in a corona around an accreting black hole. We model the spectral energy distribution of high-energy flares in this scenario. We consider particle interactions with magnetic, photon and matter fields in the black hole magnetosphere. Transport equations are solved for all species of particles and the electromagnetic output is predicted. Photon annihilation is taken into account for the case of systems with early-type donor stars.}

\FullConference{25th Texas Symposium on Relativistic Astrophysics - TEXAS 2010\\
		December 06-10, 2010\\
		Heidelberg, Germany}

\begin{document}

\section{Introduction}

The existence of a broad-band X-ray/soft gamma-ray spectra of accreting black holes in binary systems strongly suggest the presence of a very hot plasma (corona) around the central object. The soft X-ray photons produced in the inner accretion disk are Comptonized by the corona \cite{shakura} producing a power-law, high-energy feature in the spectrum.

The effects of a non-thermal population of electrons in a hot corona were considered by Kunusose \& Mineshige \cite{kunusose} and more recently by Belmont et al. \cite{belmont} and Vurm \& Poutanen \cite{vurm}. Romero et al. \cite{romero01} have studied the case of a steady state magnetized corona with injection of both relativistic electrons and protons. In the present work we extend the latter work to time-dependent injection and calculate the expected electromagnetic emission during a transient event in a hot magnetized corona around a galactic black hole. We also study the absorbing effect of an anisotropic photon-field, provided by the companion star. 

This paper is motivated by the recent detections of very high energy emission from several X-ray binaries. For example, the detection of the well known source Cygnus X-1 during a flare episode constitutes the first putative evidence of very high-energy gamma-ray emission produced around galactic black hole \cite{paredes}. More recently, the flaring nature of Cyg X-1 in gamma rays has been confirmed with the AGILE satellite \cite{sabatini}. In addition, four gamma-ray flares were detected by AGILE satellite from the X-ray binary Cygnus X-3 \cite{tavani}. The Fermi Large Area Telescope (LAT) has also detected a variable high-energy source coinciding with the position of Cyg X-3 \cite{abdo}. 

Here, we apply our model to the X-ray binary Cyg X-1.

\section{Steady state model}

We consider a spherical corona with a radius $R_{\rm{c}}=35 R_{\rm{g}}$, and a luminosity of 1 \% of the Eddington luminosity of a $10 M_{\odot}$
black hole. Since the corona is in steady state, by considering equipartition of energy we can estimate the values of magnetic field and plasma density, which result in $B = 5.7 \times 10^5$ G and $n_{e} \sim n_{i} = 6.2\times 10^{13}$cm$^{-3}$, respectively. It is assumed that the corona is composed of a  two-temperature plasma, with an electron temperature $T_{e}=10^9$ K and an ion temperature $T_{i}=10^{12}$ K (e.g., \cite{narayana},\cite{narayanb}). The X-ray emission of the corona is characterized by a powerlaw with an exponential cut-off at high energies. 

We consider an static corona where relativistic particles can be removed by diffusion. In the Bohm regime, the diffusion coefficient is $D(E) = r_{\rm{g}}c/3$, where $r_{\rm{g}} = E/(eB)$ is the gyro-radius of the particle. 

For further details on the static corona model we refer the reader to \cite{romero01}.

\begin{figure*}[!h]
\centering
\subfigure[Electron losses.]{\label{fig:perdidas:a}\includegraphics[width=0.45\textwidth,keepaspectratio]{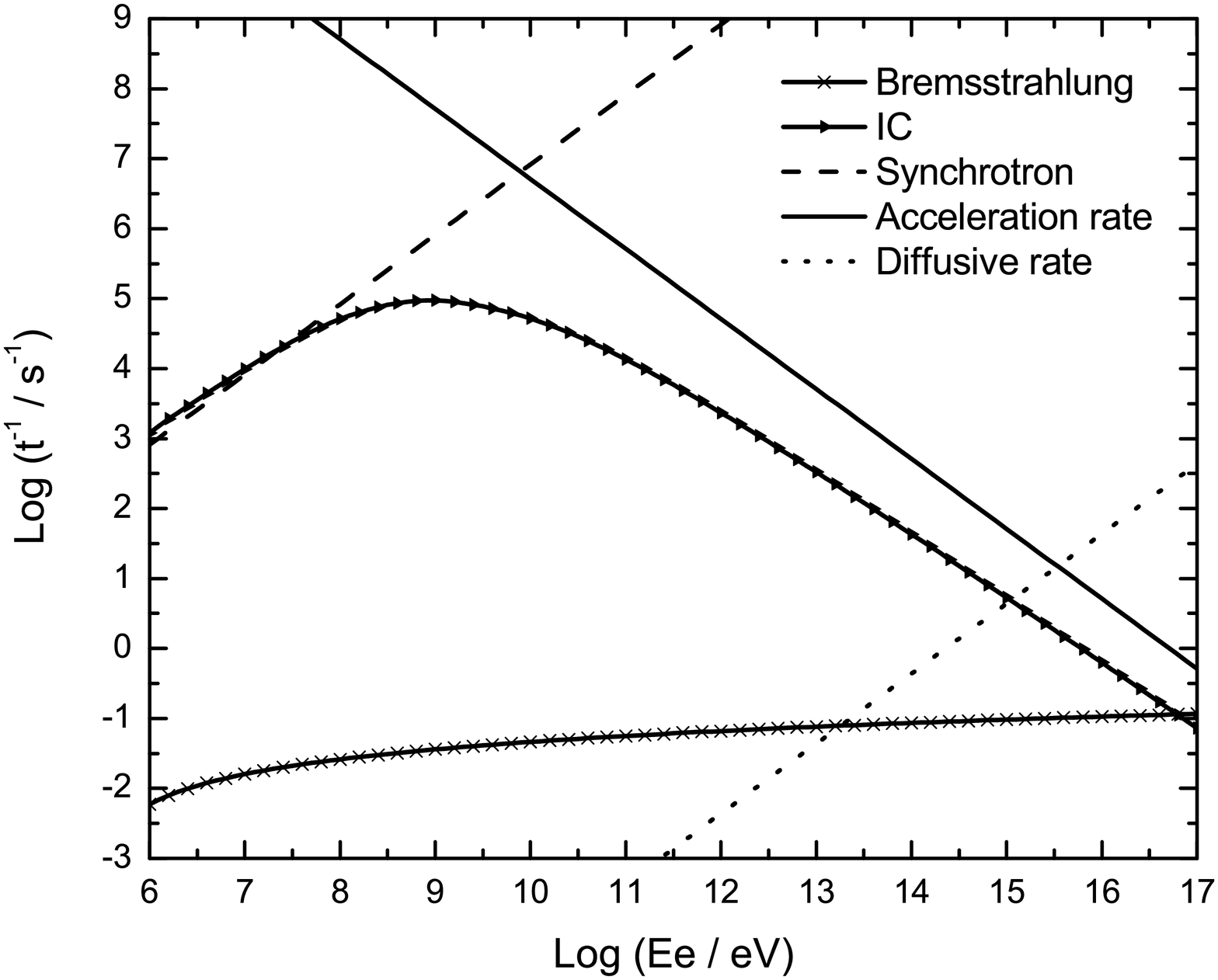}} \hspace{20pt} 
\subfigure[Proton losses.]{\label{fig:perdidas:b}\includegraphics[width=0.45\textwidth,keepaspectratio]{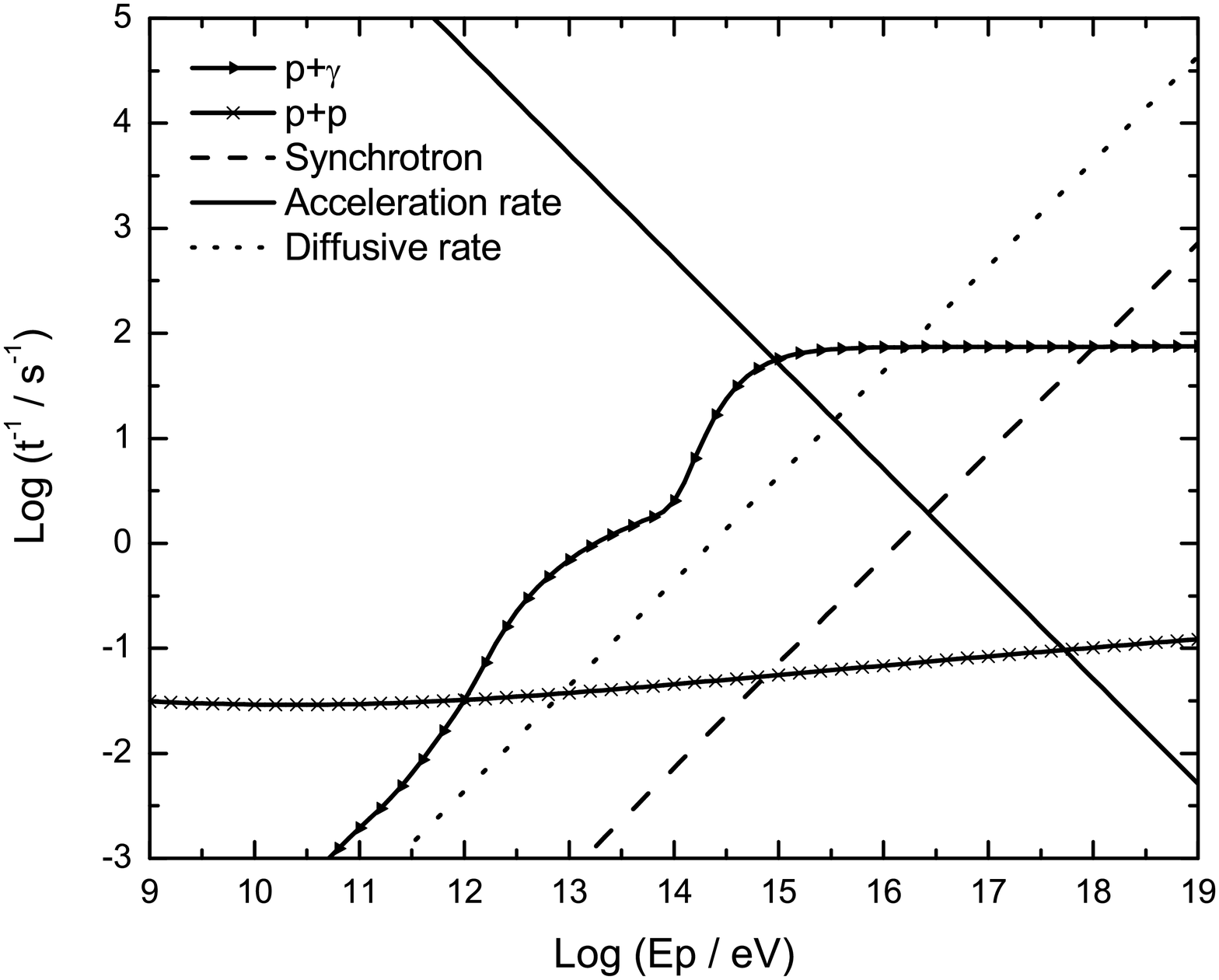}} \hfill \\ 
\caption{Radiative losses in the static corona previously described \cite{romero01}.}
\label{fig:perdidas}
\end{figure*}

In this scenario it is expected that relativistic particles lose energy mainly because of synchrotron radiation and inverse
Compton scattering, for electrons and muons, whereas for protons and charged pion the relevant cooling processes are synchrotron radiation, inellastic proton-proton collisions, and photomeson production. Figure (\ref{fig:perdidas}) shows the cooling time for different radiative processes together with the diffusion rate for primary particles.

\subsection{Spectral energy distribution}

We study the effect of the injection of a power-law distribution of relativistic electrons and protons. The main products of hadronic interactions are charged pions, which then decay producing muons and neutrinos. Neutral pions yield gamma-rays, that are a source of secondary pairs. Therefore, we also include the effect of the presence of all these secondary particles in our treatment. We solve the transport equation in steady state obtaining particle distributions for the different species.

\subsubsection{Internal absorption}

\begin{figure}[!h]
\centering
\includegraphics[clip,width=0.5\textwidth, keepaspectratio]{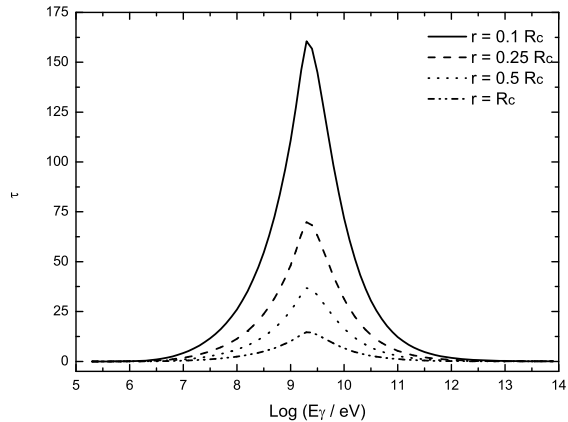}
\caption{Internal absorption due to photon-photon pair production.}
\label{fig:opacity}
\end{figure}

Gamma-rays produced in the corona can be absorbed by different mechanisms. The most relevant mechanism in this scenario is photon-photon annihilation. The absorption can be quantified by the absorption coefficient or opacity $\tau$. In Fig. (\ref{fig:opacity}) we show the opacity due to the interaction between gamma-rays and thermal X-ray photons from the corona, which are by far the dominant electromagnetic component. 

\begin{figure}[!h]
\centering
\includegraphics[clip,width=0.7\textwidth, keepaspectratio]{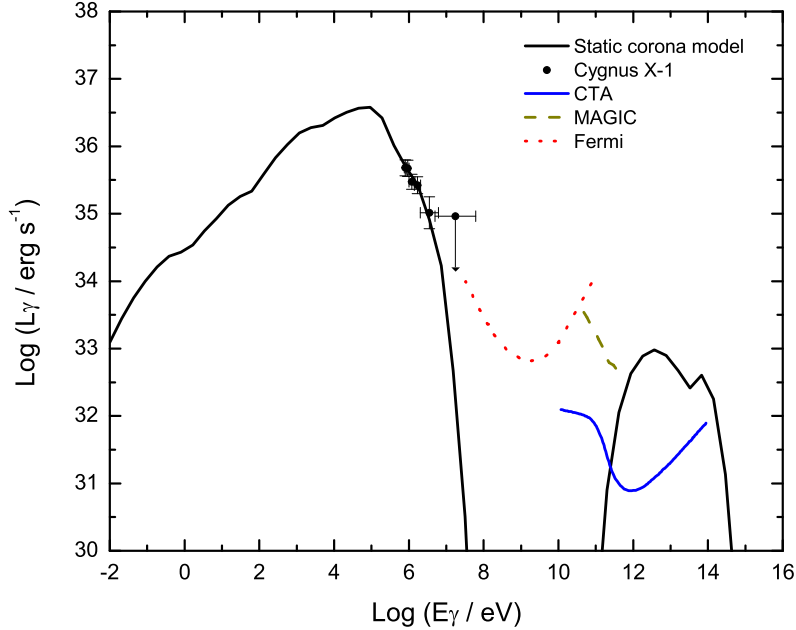}
\caption{Spectral energy distribution obtained with the steady state model of a static corona. The prediction fits the observation made by COMPTEL and {\it INTEGRAL} of Cygnus X-1 (\cite{McConnell,cadolle}).}
\label{fig:ajuste}
\end{figure}

Because of the high values of the opacity, it is expected a great number of secondary pairs to be created. The effects of internal absorption and the radiation emitted by secondary pairs are also include in the final SED, which is shown in Fig. (\ref{fig:ajuste}). This figure also shows the spectrum of the well-known source Cygnus X-1, detected by COMPTEL \cite{McConnell} and INTEGRAL \cite{cadolle}.

\subsubsection{Gamma-ray opacity due to photopair-production in the stellar radiation field}

The binary system Cygnus X-1 is composed by a massive star and a compact object. The massive star is an O9.7 Iab of $40 \pm 10 M_{\odot}$ \cite{ziolkowski}. The orbit of the system is circular, with a period of 5.6 days and an inclination between $25^{\circ}$ and $65^{\circ}$ \cite{gies}.

\begin{figure}[!h]
\centering
\includegraphics[clip,width=0.9\textwidth, keepaspectratio]{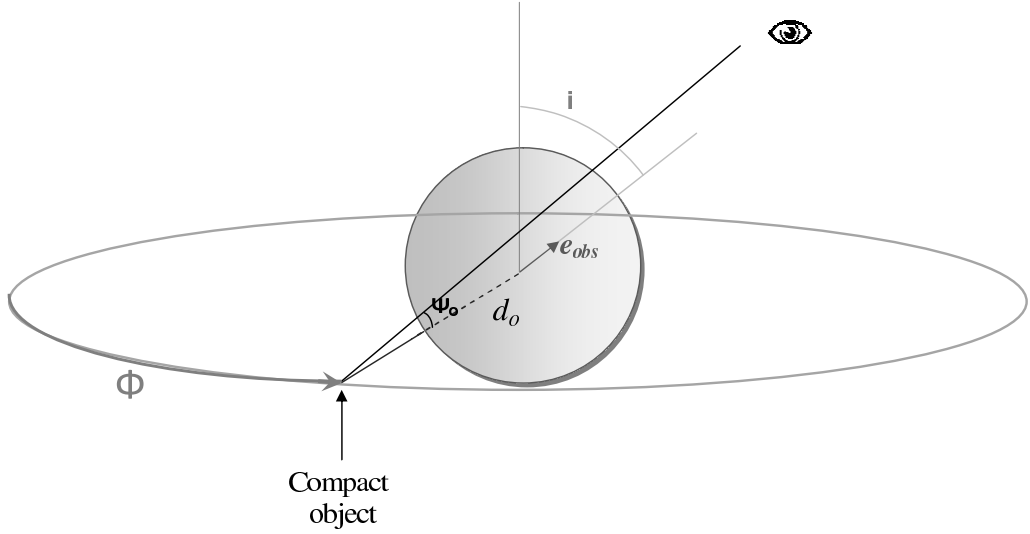}
\caption{Sketch of the geometry considered for the gamma-ray absorption in the stellar photon field \cite{romero02}.}
\label{fig:esquema}
\end{figure}

The star provides an intense radiation field that can absorb gamma-rays by pair creation within the binary system. The photon field provided by the star is anisotropic, because it depends on the position of the black hole in its orbit (see Fig. \ref{fig:esquema}). We consider the opacity treatment for gamma-ray absorption in Cygnus X-1 as in \cite{romero02}.

The star has a radius $R_{*}=1.5 \times 10^{12}$ cm, and for simplicity we asume a blackbody density radiation of a temperature $T_{*}=3 \times 10^{4}$ K. The orbital radius is $r_{\rm{orb}} = 3.4 \times 10^{12}$ cm. The value of the orbital phase $\phi=0$ corresponds to the compact object at opposition (Fig. \ref{fig:esquema}).

\begin{figure}[!h]
\centering
\includegraphics[clip,width=0.5\textwidth, keepaspectratio]{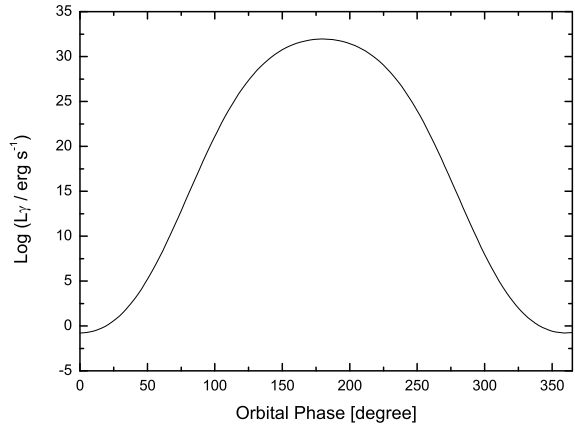}
\caption{Modulation of gamma-ray emission in steady state at $E \sim 50$ GeV by the anisotropic photon field of the companion star.}
\label{fig:modulation}
\end{figure}

In Fig. (\ref{fig:modulation}) we show the gamma-ray emission of the corona in steady state at a given energy. The modulation is due to the photon absorption in the stellar field.

\section{Flare model}

The temporal dependence of the particle injection is characterized by a FRED (Fast Rise and Exponential Decay) behavior, whereas the energy dependence is a power-law. The transient injection can be represented by \cite{reynoso}

\begin{equation}
Q(E,t) = Q_{0} E^{-\alpha} e^{-E/E_{\rm{max}}} (1-e^{t/\tau_{\rm{rise}}} ) \left[ \frac{\pi}{2}- \arctan \Big( \frac{t-\tau_{\rm{plat}}}{\tau_{\rm{dec}}} \Big) \right],
\end{equation}

\noindent where $\tau_{\rm{rise}} = 30 ^{\rm{min}}$, $\tau_{\rm{dec}} = 1^{\rm{h}}$ and $\tau_{\rm{plat}} = 2^{\rm{h}}$. The power-law has the standard index of $\alpha=2.2$. The normalization constant $Q_{0}$ can be obtained from the total power injected in relativistic protons and electrons, $L_{\rm{rel}}=L_{p}+L_{e}$. This power is assumed to be a fraction of the luminosity of the corona, $L_{\rm{rel}}=q_{\rm rel} L_{\rm{c}}$. In the steady state the best fit to the observations is obtained with $q_{\rm rel}=0.2$. During the flare the number of relativistic particles increases. In our model, the power injected in the flare doubles of that injected in steady state. 

In order to obtain the evolution of particle distributions $N(E,t)$ for each type of particle, we solve the transport equation given by \cite{ginzburg}

\begin{equation}\label{eq:transporte}
		\frac{\partial N(E,t)}{\partial t} +  \frac{\partial }{\partial E} \Big( b(E) N(E,t) \Big)+ \frac{N(E,t)}{t_{\rm{esc}}}=Q(E,t) .
	\end{equation} 

\noindent where $b(E)= \frac{dE}{dt} \Big | _{\rm{loss}}$. 

\begin{figure}[!h]
\centering
\includegraphics[clip,width=0.7\textwidth, keepaspectratio]{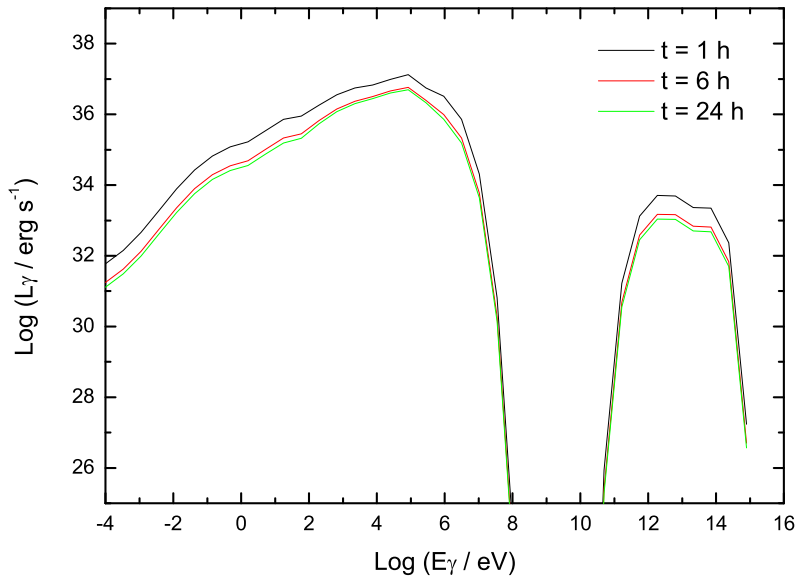}
\caption{Evolution of the luminosity during the flare, without consider the absorption due to the star.}
\label{fig:SEDevolution}
\end{figure}

\begin{figure}[!ht]
\centering
\subfigure[$\phi_{0} = 0.0$]{\label{fig:tauExterna:a}\includegraphics[width=0.45\textwidth,keepaspectratio]{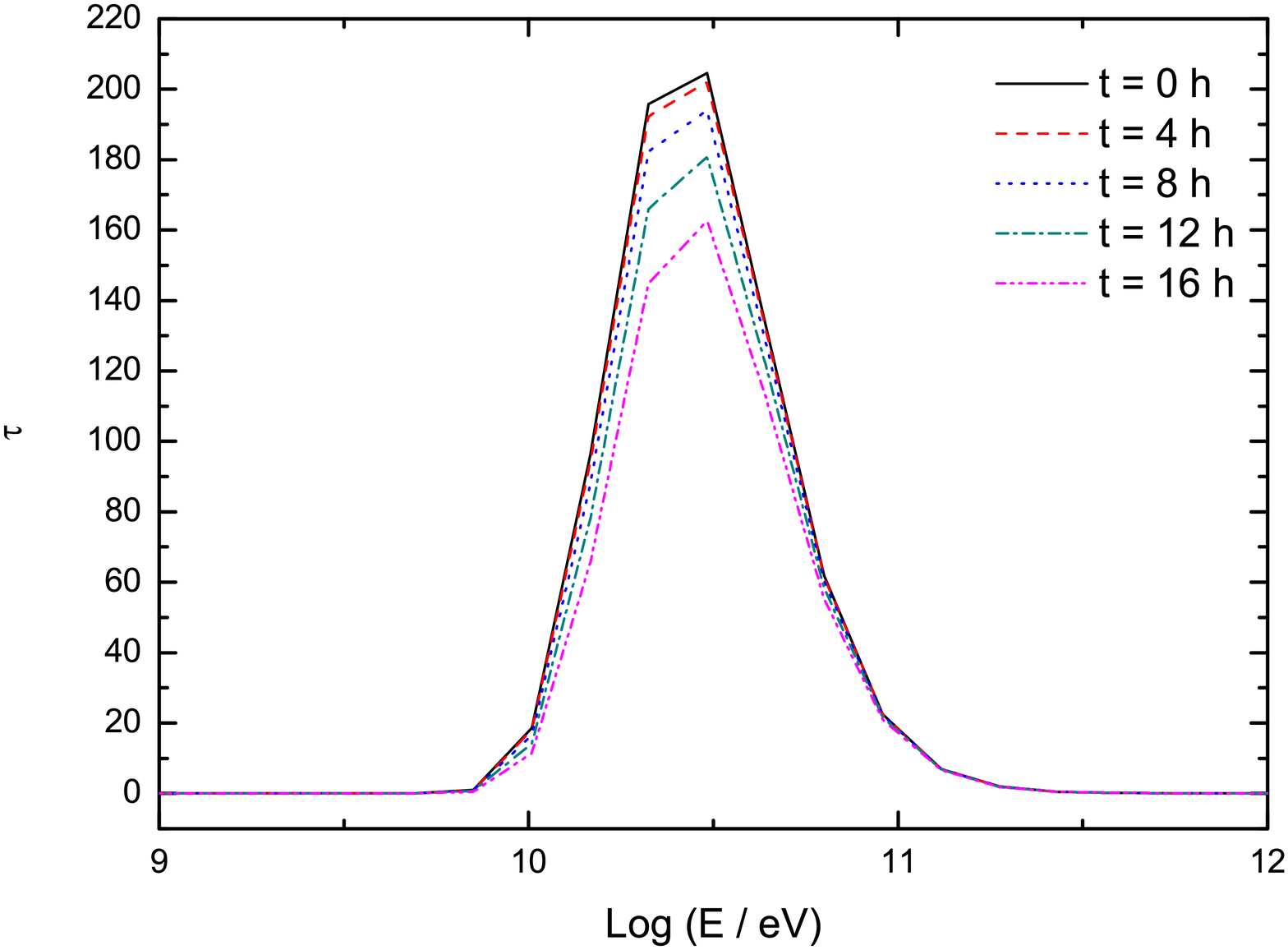}} \hspace{20pt} 
\subfigure[$\phi_{0} = \pi/2$]{\label{fig:tauExterna:b}\includegraphics[width=0.45\textwidth,keepaspectratio]{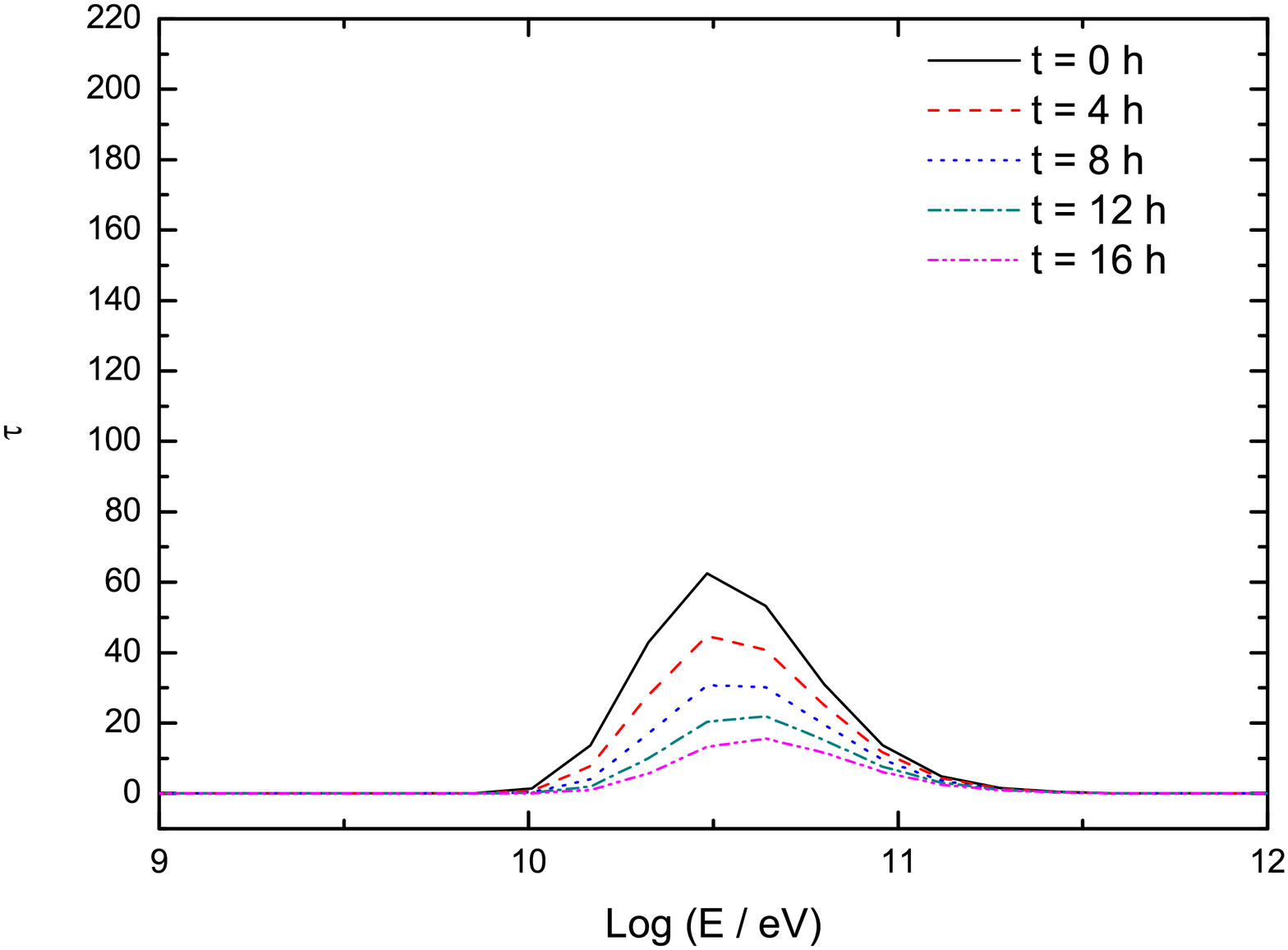}} \hfill \\ 
\subfigure[$\phi_{0} = \pi$]{\label{fig:tauExterna:c}\includegraphics[width=0.45\textwidth, keepaspectratio]{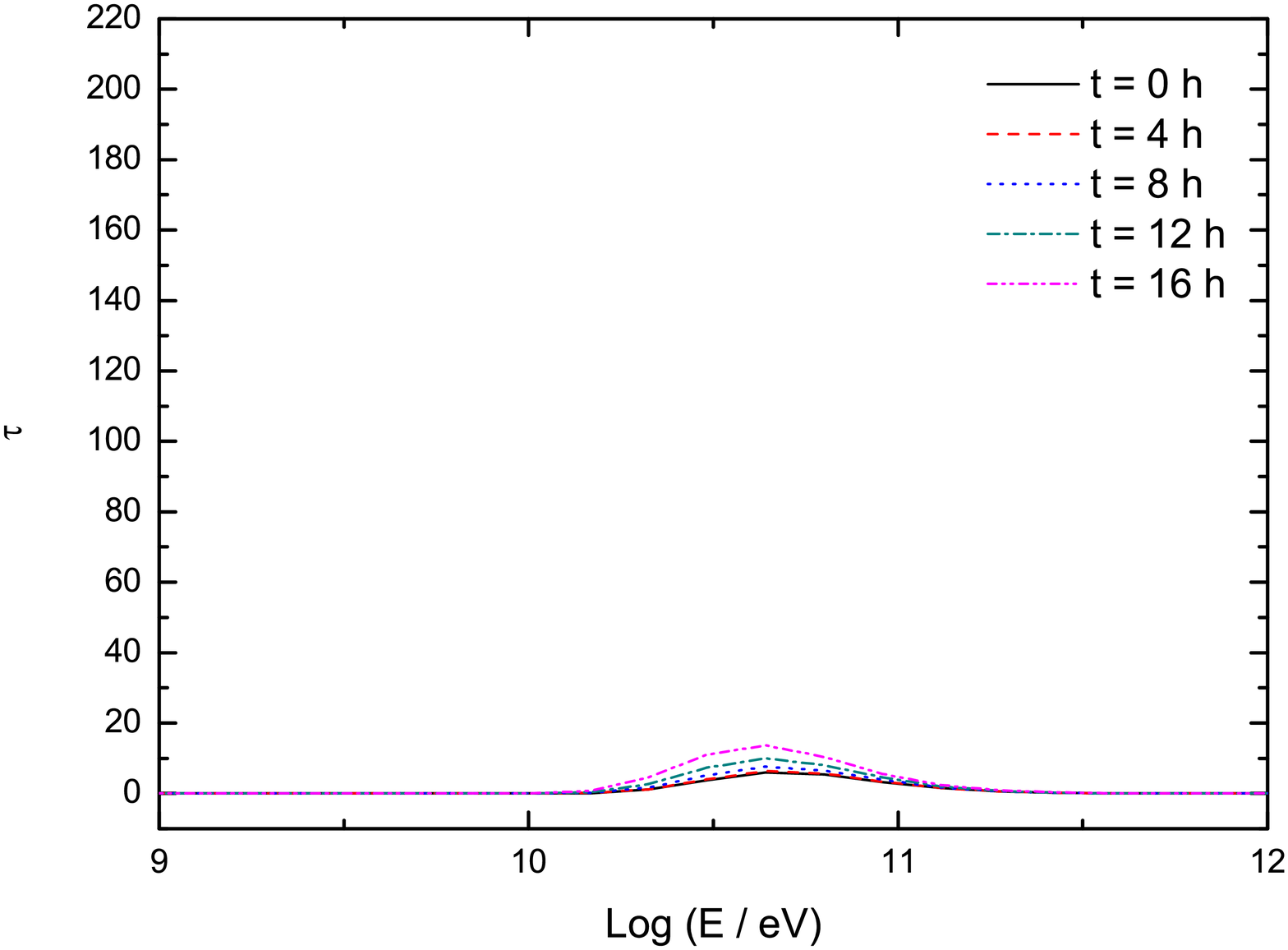}}  \hspace{20pt}
\subfigure[$\phi_{0} = 3\pi/2$]{\label{fig:tauExterna:d}\includegraphics[width=0.45\textwidth, keepaspectratio]{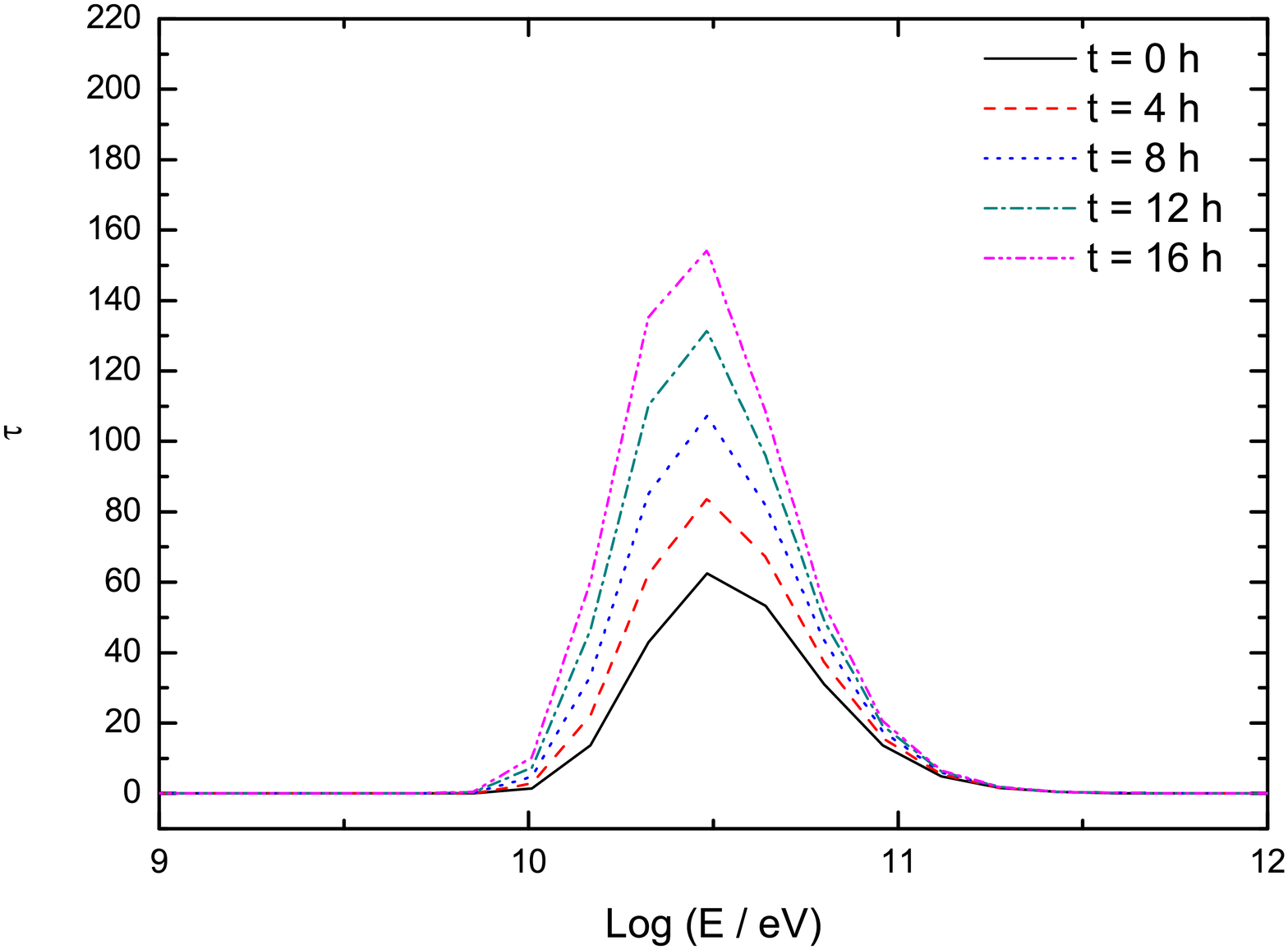}} \hfill
\caption{Opacity due to the presence of an anisotropic photon field.}
\label{fig:tauExterna}
\end{figure}

Fig. (\ref{fig:SEDevolution}) shows the evolution of the electromagnetic emission during a day, not corrected by the absorption in the photon field of the star. For this purpose, in Fig. (\ref{fig:tauExterna}) we show the absorption coefficient due to the stellar field. We estimate the opacity for flares occurring at different orbital phases, and we conclude that for some values of $\phi$ the absorption of the star is almost negligible.

\section{Summary and conclusions}

We have estimated the electromagnetic output produced in the corona during a transient outburst. Our study is based on a steady state model of an accreting black hole, which is in agreement with the observations of Cygnus X-1 \cite{romero01}.  We consider both internal and external absorption. We have included the effect of the presence of an anisotropic stellar photon-field. The absorption produced by this field results almost negligible for certain positions of the compact objects; nevertheless it affects the electromagnetic emission at $E \sim 30$ GeV for the most of the orbit.

In a future work, we will study the modulation in the emission for objects in which the orbital period is comparable with the flare duration.


\end{document}